\documentclass[12pt]{article}
\usepackage{graphicx}
\begin{document}

\def\hh{ \hat h }
\def\nro{ n_{R_1} } 
\def\nrt{ n_{R_2} } 
\def\nrl{ n_{R_l} } 
\def\nso{ n_{S_1} } 
\def\nst{ n_{S_2} } 
\def\nsk{ n_{S_k} } 
\def\nri{ n_{R_i} } 
\def\nsj{ n_{S_j} } 
\def\rop{ R_1^{\prime} } 
\def\rtp{ R_2^{\prime} } 
\def\sop{ S_1^{\prime} } 
\def\stp{ S_2^{\prime} } 
\def\rlp{ R_l^{\prime} } 
\def\skp{ S_k^{\prime} } 
\def\rip{ R_i^{\prime} } 
\def\sjp{ S_j^{\prime} } 
\def\car{\cal R}
\def\eps{\epsilon}
\def\yt{{\tilde y}}
\def\at{{\tilde a}}
\def\Xt{{\tilde X}}
\def\omegat{{\tilde \omega}}
\def\xit{{\tilde \xi}}
\def\Ft{{\tilde F}}
\def\qt{{\tilde q}}
\def\CN{{\cal N}}
\def\mt{{\tilde m}}
\def\Yt{{\tilde Y}}
\def\CO{{\cal O}}
\def\CF{{\cal F}}

%
%
\def\PTP{ {\it Prog. Theor. Phys.}}
\def\NP{{\it Nucl. Phys.\ }}
\def\AP{{\it Ann. Phys.\ }}
\def\PL{{\it Phys. Lett.\ }}
\def\PR{{\it Phys. Rev.\ }}
\def\PRL{{\it Phys. Rev. Lett.\ }}
\def\CMP{{\it Comm. Math. Phys.\ }}
\def\JMP{{\it J. Math. Phys.\ }}
\def\JSP{{\it J. Stat. Phys.\ }}
\def\JTP{{\it JETP \ }}
\def\JTPL{{\it JETP Lett.\ }}
\def\JP{{\it J. Phys.\ }}
\def\IJMP{{\it Int. Jour. Mod. Phys.\ }}
\def\Mod{{\it Mod. Phys. Lett.\ }}
\def\NC{{\it Nuovo Cimento \ }}
\def\PRep{{\it Phys. Rep.\ }}
\def\JGP{{\it J. Geom. Phys.\ }}
\def\ZP{ {\it Z. Phys.\ }}
\def\JHEP{{\it JHEP \ }}
\def\ATMP{{\it Adv. Theor. Math. Phys.\ }}
\def\CQG{{\it Class. Quantum Grav.\ }}

\begin{titlepage}
\thispagestyle{empty}
\begin{flushright}
BROWN-HET-1313
\end{flushright}
\vskip 1cm
\begin{center}
{\LARGE\bf Probing the Enhancon via Calculations in
  Supersymmetric Gauge Theory}
\vskip 1cm

{\large Steven Corley and David A. Lowe}
\vskip .5cm
{\it Department of Physics \\
Brown  University \\
Providence, RI 02912}\\
{\tt scorley, lowe@het.brown.edu}

\end{center}
\vskip 1cm

\begin{abstract}
We consider the $\CN=2$ gauge theory on $N$ D7-branes wrapping $K3$, with
D3-brane probes. In the large $N$ limit, the D7-branes blow up to form
an enhancon shell. We probe the region inside and outside the enhancon shell using
the D3-branes, and compute the probe metric using the
Seiberg-Witten formalism. Supergravity arguments suggest a flat
interior up to $1/N$ corrections, and indeed our results for
the D3-brane probes are consistent with that.
By including the dynamics of the branes, these results, together
with those of hep-th/0204050, demonstrate the robustness of the
enhancon 
mechanism
beyond patching together of supergravity solutions with D-brane
source junction conditions.

\end{abstract}

\end{titlepage}

\section{Introduction}

One of the most important questions that a consistent theory of
quantum gravity should answer is how spacetime singularities are to be
resolved. This is of particular importance in cosmology, where we
believe the big bang emerged from a spacelike or null singularity. In
the present paper we will consider the resolution of time-like
singularities in string theory using the so-called enhancon
mechanism. Such singularities are presently under far better
calculation control than the singularities of interest in cosmology,
because one may use techniques from supersymmetric gauge theory.

The enhancon mechanism is the way string theory resolves a particular
type of timelike singularity associated with a stack of D-branes
wrapping a compact cycle of the internal compactified 
space \cite{Johnson:1999qt}. The
would-be timelike singularity of the naive supergravity solution is
resolved by brane sources expanding out to form a shell. Inside the
shell of brane sources, it has been argued the spacetime is flat. 

The main motivation for the enhancon picture comes from studying the
behavior of test brane probes. In \cite{Johnson:1999qt}, a
configuration of coincident D-branes wrapping a $K3$ surface was considered, and
this was probed by a test D-brane also wrapping the $K3$.
The coefficient of the
$v^2$ term in the effective action for the probe
changes 
sign at
a point of order a string length away from a would-be naked
singularity in the naive supergravity solution, where the volume of
the $K3$ reduces to the self-dual point. This coefficient of the $v^2$
term can
be thought of as inertial tension. The change in
sign of the inertial tension suggests the following self-consistent
picture: the wrapped D-branes expand out to form an enhancon shell,
and incoming probe D-branes spread out as they approach the shell and
dissolve into it as their inertial tension vanishes. In particular, 
the region inside the enhancon shell cannot be probed by wrapped
D-branes. In earlier work \cite{Alberghi:2002tu}, 
we computed the moduli space metric for
such wrapped probe branes for the case of D7-branes, 
by performing calculations using the low
energy effective action of the $\CN=2$ supersymmetric $SU(N)$ gauge
theory. This low energy description of the dynamics is appropriate
when the probe branes are within a string length of the enhancon shell.
At large $N$, we found agreement with supergravity predictions, but we also
obtained $1/N$ corrections, as well as instanton corrections
nonperturbative in $1/N$. See
\cite{Ferrari:2001mg,Ferrari:2001zb} for other studies of the enhancon
from the gauge theory perspective.

An unwrapped D-brane probe has a rather different experience as it
approaches the enhancon shell. For the remainder of this paper we will
restrict our attention to an enhancon configuration built out of
D7-branes wrapping a $K3$, and will add unwrapped D3-branes as probes.
Since the D3-brane does not wrap the $K3$, its
inertial tension remains non-zero as it passes through the enhancon
shell \cite{Johnson:2001wm}. The main focus of the present paper will
be to  use such probes to examine the interior of the
enhancon shell. We will  solve for the low-energy effective action of
such probes using techniques from $\CN=2$ supersymmetric gauge theory 
\cite{Seiberg:1994rs,Seiberg:1994aj}.

These interior probes are interesting, because the region inside need not be governed by
low energy supergravity. Of course if one assumes the low energy
supergravity equations are valid in the interior, flat space is guaranteed by
Gauss' law. 
However this assumption need not be correct a priori. From
the point of view of solving the low energy equations of motion, with
boundary conditions placed at spatial infinity, the interior of the
enhancon is inside a shell where stringy physics is important since the
curvature becomes of order the string scale near the shell. So the
possibility that stringy physics is important in the interior cannot
be ruled out without further consideration. We will find that curvature in
the interior is actually of order $N^{-2}$ due to back-reaction effects.

Furthermore the patching of supergravity solutions
considered in \cite{Johnson:1999qt,Johnson:2001wm} relies  heavily on the
presence of 
unbroken supersymmetry, so it is conceivable the solution is unstable to generic
perturbations. The enhancon shell is the minimal radius at which D-brane
source boundary conditions can be consistently placed, but any larger
radius will also do, so the configuration is at best marginally stable
\cite{Johnson:2001wm}. 
By exactly solving the low-energy effective
action for probe branes in this background, we are able to take into
account effects of interactions and small deviations away from
supersymmetry. Our results place the enhancon mechanism on a much more
robust footing. 

At the same time, the gauge theory results also highlight the
limitations of the enhancon mechanism. Once one gives up spherical
symmetry
in the large $N$ limit, one can try to consider spacetime
singularities dual to Argyres-Douglas \cite{Argyres:1995jj} 
points in the moduli space of
the gauge theory. At these points an enhancon mechanism does not
appear to work \cite{Alberghi:2002tu}. 
It remains an interesting open problem to understand the
spacetime physics associated with these singularities.

A summary of the layout of the paper is as follows. In section 2 we
review the supergravity solution for D7-branes wrapping $K3$ and the
enhancon mechanism in this context. In section 3 we probe the enhancon
using a pair of unwrapped D3-branes by solving for the low energy
effective action of $\CN=2$
supersymmetric $SU(N)$ gauge theory with two fundamental
hypermultiplets. In section 4 we briefly consider the case
of a single hypermultiplet, which is technically more complicated. In
section 5, we use these results to compute the moduli space metric for
the probe, and compare to
supergravity expectations.
Section 6 contains some conclusions and prospects for the future.

\section{Supergravity solution}

We consider the supergravity solution for $N$ D7-branes wrapping $K3$
\cite{Johnson:1999qt}.
The spacetime solution in string metric is
\begin{eqnarray}
\label{sugrasol}
ds^2 & = & Z_3^{-1/2} Z_7^{-1/2} \eta_{\mu \nu} dx^\mu dx^\nu +
Z_3^{1/2} Z_7^{1/2} (\alpha')^2 du d \bar u + (2 \pi R)^2 Z_3^{1/2}
Z_7^{-1/2} ds_{K3}^2 \\
e^\Phi &=& g Z_7^{-1} \nonumber \\
Z_3 &=& {g N \over 2 \pi} { (\alpha')^2 \over R^4} \ln( U /\rho_3)
\nonumber \\
Z_7 & = & {g N \over 2 \pi} \ln( \rho_7 /U) ~,\nonumber
\end{eqnarray}
where $U=|u|$, $V=(2\pi R)^4$ is the $K3$ volume and $g$ is the string
coupling. Here $\eta_{\mu\nu}$ is the Minkowski metric, and
$ds_{K3}^2$ is the $K3$ metric.
The constants $\rho_3$
and $\rho_7$ are new constants of integration that appear in the
supergravity solution. For $N\leq 24$ this type of configuration can
be realized in F-theory. However for $N>24$ this supergravity solution
must be embedded in some yet to be discovered
generalization of F-theory.
This solution has a singularity at $U=\rho_7$ where the dilaton blows
up and the supergravity description breaks down. $U=\rho_3$ is the
repulson singularity (we assume $\rho_3 < U < \rho_7$), where the
dilaton remains finite, but the curvature blows up. This singularity
is cutoff by the enhancon mechanism, with the enhancon shell sitting
just outside $U=\rho_3$.

For a wrapped D7-brane probe, the inertial tension is 
\begin{equation}
\label{biaction7}
{1\over 2 g} (\mu_7 V Z_3 - \mu_3 Z_7)~,
\end{equation}
where $\mu_3 = (2 \pi)^{-3} {\alpha'}^{-2}$ and 
$\mu_7 = (2 \pi)^{-7} {\alpha'}^{-4}$. This changes sign as discussed
above, as $U$ approaches $\rho_3$. This is the sign of the instability
leading to the expansion of the brane charge into the enhancon
shell. The inertial tension vanishes at the enhancon radius. Because
the Compton wavelength of these branes becomes large near the enhancon
shell, we cannot use these branes as probes of the interior of the enhancon.

For an unwrapped D3-brane, the inertial tension is
\begin{equation}
\label{biaction3}
{1\over 2 g}  \mu_3 Z_7~,
\end{equation}
As remarked above, this is finite at the enhancon radius $U=U_e$, and remains
finite inside the enhancon shell, where $Z_7$ is to be replaced by
$Z_7(U_e)$. These branes can be used as probes of the interior of the
enhancon shell, where they simply see a flat moduli space metric.

\section{Probing the enhancon}
\label{sec:twohyper}

We begin by considering the case of two D3-brane probes in the limit
that they approach the collection of D7-branes. As we will see,
considering two probe branes
leads to technical simplifications that allow us to obtain
explicit results for the period matrix. We will work in a
limit where the inertial tension of the D3-branes is much smaller than that of
the $N$ wrapped D7-branes so that they may be considered true test probes of
the enhancon geometry. In this limit the dynamics reduces to simply $SU(N)$
with two fundamental hypermultiplets, rather than the full 
$SU(N)\times  U(2)$ gauge theory with a
bifundamental  $(N,2)$ hypermultiplet. We will comment further on this
decoupling later.
As is now
well-known, the low energy effective action can be obtained from the
following Riemann curve \cite{Hanany:1995na,Argyres:1995xh,Klemm:1995qs}
\begin{equation}
\label{twohypercurve}
Y^2 = A(X)^2-B(X)
\end{equation}
with
$A(X)=\prod_{i=1}^N (X-\phi_i)$ and 
$B(X)= \Lambda^{2N-n_3}\prod_{f=1}^{n_3} (X+m_f)$
where the QCD scale is $\Lambda$ and for us $n_3=2$ with $m_f$  the masses of each
hypermultiplet, and the $\phi_i$ are the classical moduli space
coordinates corresponding to the transverse positions of the D7-branes.

For simplicity,
we take all of the classical moduli space coordinates $\phi_i=0$ 
as this is the point that corresponds to the spherically
symmetric enhancon
\cite{Johnson:1999qt,Alberghi:2002tu}. 
To investigate the physics inside the enhancon we take the
masses of the hypermultiplets to be small, i.e.,
$m_{1}/\Lambda,m_{2}/\Lambda \ll 1$.  Naively this is
the correct limit to take as these masses measure the
length of the strings stretched between the constituent
D7-branes and the probe D3-branes.
The branch points separate into two classes.
Most ($2N-2$) of the branch points lie on a circle (the enhancon) 
about the origin.  In a double expansion in $m_1/\Lambda$ and
$m_2/\Lambda$ these branch points are given by
\begin{eqnarray}
X_{k} & = & X_{k}^{(0)} (1 + X_{k}^{(1)} + X_{k}^{(2)}+ \cdots) \\
X_{k}^{(0)} & = & \Lambda e^{i \pi (k-1)/(N-1)} \nonumber \\
X_{k}^{(1)} & = & \frac{\Lambda^{2N-2} (m_1 + m_2) }
{2 N (X_{k}^{(0)})^{2N-1} - 2 \Lambda^{2N-2} ~ X_{k}^{(0)}} \nonumber \\
X_{k}^{(2)} & = & \frac{\Lambda^{2N-2} ( (X_{k}^{(0)} X_{k}^{(1)})^{2}
+ (m_1 + m_2) X_{k}^{(0)} X_{k}^{(1)} + m_1 m_2) - N (2N-1) 
(X_{k}^{(0)})^{2N} (X_{k}^{(1)})^2}{2 N (X_{k}^{(0)})^{2N}
- 2 \Lambda^{2N-2} (X_{k}^{(0)})^{2}} \nonumber
\label{twoenhroots}
\end{eqnarray}
for $1 \leq k \leq 2(N-1)$.
The remaining two roots lie near the center of the circle and
are given approximately by
\begin{eqnarray}
X_{m_1} & = & -m_1 \Big( 1 - \frac{1}{2} (1-\frac{m_2}{m_1}) - \frac{1}{2}
\sqrt{(1-\frac{m_2}{m_1})^2 + 4 (\frac{m_1}{\Lambda})^{2(N-1)}}+
\cdots \Big) \nonumber \\
X_{m_2} & = & -m_2 \Big( 1 - \frac{1}{2} (1-\frac{m_1}{m_2}) + \frac{1}{2}
\sqrt{(1-\frac{m_1}{m_2})^2 + 4 (\frac{m_2}{\Lambda})^{2(N-1)}}+
\cdots \Big)
\end{eqnarray}
We have assumed that $m_1 \leq m_2$ in writing the
above form of the roots.  Moreover
when $m_1 = m_2$ we shall always take the root
$X_{m_1}$ to satisfy $|X_{m_1}| < |X_{m_2}|$ (which
is the case if the limit is taken on the expressions
given).

To evaluate the periods we take the following basis
of cycles.  The $\alpha_k$ cycles encircle the $X_{2k}$
and $X_{2k+1}$ branch points for $1 \leq k \leq (N-2)$
and the $\alpha_{N-1}$ cycle encircles $X_{1}$ and
$X_{m_1}$, the $\gamma_j$ cycles
encircle the $X_{2j-1}$ and $X_{2j}$ branch points
for $1 \leq j \leq (N-1)$, and the $\beta_j$ cycles
are defined by
\begin{eqnarray}
\beta_j & \equiv & \gamma_N +
\sum_{k=1}^{j} \gamma_k = - \sum_{k=j+1}^{N-1} \gamma_k, ~~
1 \leq j \leq (N-2) \nonumber \\
& \equiv & \gamma_N =  -  \sum_{k=1}^{N-1} \gamma_k, ~~
j = (N-1).
\end{eqnarray}
One needs to be careful in deforming contours as we have
in the above expressions.  We will comment shortly on
why this is valid.
One can easily check that the $\alpha$ and $\beta$
cycles form a canonical basis.

\begin{figure}
\centering
\includegraphics[height=7cm]{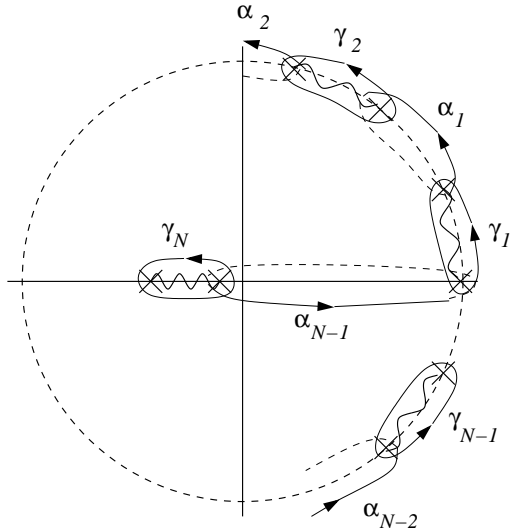}
\caption{Diagram of the branch points and cycles for the
two hypermultiplet case.}
\label{twohyper}
\end{figure}

Before continuing it is useful to discuss in a little
more detail figure \ref{twohyper}.  Roughly speaking
a pair of branch points corresponds to a D7-brane.
One has to be a little careful with such statements,
especially in the context here where we are considering
a large number of branch points ($2N$ with large $N$)
that are closely packed together, and moreover because
a given pair of branch points could actually correspond
to more than just a single D7-brane.  At any rate,
given this identification, figure \ref{twohyper}
shows that there is a single D7-brane lying near the center
of the enhancon.  What has happened is the following.
Consider first the case where the hypermultiplet masses
are large, $m_1 /\Lambda, m_2 /\Lambda \gg 1$.  In this
limit the branch points of the Seiberg-Witten curve
are given approximately by $X_k = (\Lambda^{2(N-1)} m_1 m_2)^{1/2N}
e^{i \pi k/N}$ for $1 \leq k \leq 2N$, i.e., all branch
points are on the enhancon circle so that all D7-branes
are on the enhancon circle.  In this limit one could
integrate out the hypermultiplets to arrive at the pure
gauge theory at the enhancon point of the moduli space.  
As the masses of the
hypermultiplets are decreased, the size of the enhancon
shrinks and moreover a pair of branch points moves 
toward the center of the circle.  In the limit
that both masses tend to zero, these two branch
points merge at the origin and the low-energy effective
action breaks down.  What we will now show is that
in the limit where both masses are taken to zero,
certain charged states associated to fundamental
hypermultiplets, and corresponding to the
$\beta_{N-1}$ (or $\gamma_N$) contour, are becoming massless. 
In other words, the D7-brane that lies inside the
enhancon is really a bound state of the D7 and D3-branes. 

The periods are given by integrating the Seiberg-Witten
form $d\lambda$ around the basis of contours described
above and depicted in figure \ref{twohyper}.  Specifically
one has
\begin{eqnarray}
a_j & = & \frac{1}{2 \pi i} \oint_{\beta_j} d\lambda \nonumber \\
a_{D_k} & = & \frac{1}{2 \pi i} \oint_{\alpha_k} d\lambda
\label{periodsdefine}
\end{eqnarray}
where $d\lambda$ is given by 
\cite{Klemm:1995qs,Argyres:1995xh,Hanany:1995na}
\begin{equation}
d\lambda =  \frac{X dX}{Y} 
\left({\cal C}'(X) - {\cal C}(X) \frac{G'(X)}{2 G(X)} \right) ~,
\label{SWform}
\end{equation}
and ${\cal C}(X) = \prod_{i=1}^{N} (X - \phi_i)$ and
$G(X) = \prod_{l=l}^{N_f} (X + m_l)$.  In our case
${\cal C}(X) = X^{N}$ and $G(X) = (X+m_1)(X+ m_2)$.
It is important to note that the $G'(X)/G(X)$ term
in $d\lambda$ has poles at $X=-m_1$ and $X=-m_2$.
We take our contours above so as to not encircle
these poles.  The poles nevertheless will contribute
to the mass formula as will be seen below.

To evaluate the periods on the enhancon it is convenient
to parametrize the integrals in terms of a new variable
$\varphi$ as
\begin{equation}
X(\varphi) = X_{k}|_{k \rightarrow k + \varphi/\pi}.
\label{Xparam}
\end{equation}
The usefulness of this parametrization lies in the fact
that $Y$ is given exactly by
\begin{equation}
Y(\varphi) = \Lambda^{N} e^{i \pi k/(N-1)} e^{i \varphi/(N-1)}
\sqrt{e^{i 2 \varphi} -1}.
\label{Yexpans}
\end{equation}
As a consequence,
substituting the expansion (\ref{Xparam}) into the Seiberg-Witten
form (\ref{SWform}) one finds that all $\varphi$ integrals reduce to the
form
\begin{eqnarray}
F(n) & = & \int_{0}^{\pi} d \varphi \frac{e^{i(n-1) \varphi/(N-1)}}
{\sqrt{e^{i 2 \varphi} -1}} = \frac{e^{-i \pi/4}}{\sqrt{2}}
\int_{0}^{\pi} d \varphi \frac{e^{i((n-1)/(N-1) -1/2) \varphi}}
{\sqrt{\sin (\varphi)}} \\
& = & -i2 \sqrt{\pi} {N-1 \over n-1} {\Gamma(1+(n-1)/(2N-2)) \over 
\Gamma(1/2 + (n-1)/(2N-2))}
e^{in \pi/(2N-2)} \sin \left( \frac{(n-1) \pi}{2N-2} \right)~. \nonumber
\label{Fexplicit}
\end{eqnarray}
This leads to the expressions for the $\at_j$ (corresponding
to the $\gamma_j$ contours) and $a_{D_k}$ periods
\begin{eqnarray}
\at_j & = &  \frac{\Lambda}{\pi (1 - 1/N)} \biggl(
e^{i 2 \pi N (j-1)/(N-1)} F(N+1) \left( 1 - \frac{1}{N} \right)
\nonumber \\ & - & \frac{m_1 + m_2}{2 \Lambda}
e^{i 2 \pi (j-1)} \left( \frac{F(-N + 2)}{N-1}
- \frac{F(N)}{N} \right) + \cdots \biggr), ~ 1 \leq j \leq (N-1)
\nonumber \\
a_{D_k} & = & \at_j |_{j \rightarrow k + 1/2}, ~ 1 \leq k \leq (N-2).
\label{periods}
\end{eqnarray}
To higher order
we show in the appendix that the expansion of the periods
in terms of the mass parameter $m$ defined as $m \equiv m_1$
with $\xi \equiv  m_2/m_1$ is given by
\begin{equation}
\at_j = \sum_{n=0}^{\infty} c_{n}(\xi) \left( \frac{m}{\Lambda}
\right)^n  e^{i 2 \pi (j-1) (1-n)/(N-1)} ~,
\label{periodseries}
\end{equation}
with a similar expansion for $a_{D_k}$.

As an application of this expansion we note that
due to the phase factor one obtains the formula
for the sum of periods
\begin{equation}
\sum_{j=1}^{N-1} \at_j = (N-1) \sum_{p=0}^{\infty}
c_{p(N-1) + 1} (\xi) \left( \frac{m}{\Lambda} \right)^{p(N-1) +1}~,
\label{aninter}
\end{equation}
where from (\ref{periods}) we find that $c_1 (\xi)$ is given by
\begin{eqnarray}
c_1 (\xi) & = & - \frac{1+ \xi}{2 \pi (1-1/N)}
\left( \frac{F(-N+2)}{(N-1)} - \frac{F(N)}{N} \right)
\nonumber \\
& = & \frac{1 + \xi}{2 (N-1)}~,
\end{eqnarray}
where we have used $F(-N+2)=0$ and $F(N)= \pi$ as follows
from (\ref{Fexplicit}).  However the sum (\ref{aninter})
corresponds to integrating $d\lambda$ around the contour
$\sum_{j=1}^{N-1} \gamma_j$, which can be deformed to
the contour $- \gamma_N$ (which by definition does not include the
poles at $X=-m_1$ and $X=-m_2$) plus the contribution of the
poles of $d\lambda$ at $X=-m_1$ and $-m_2$ plus the
contribution due to the pole at infinity.  One can show
easily however that the contribution from the pole at 
infinity exactly cancels the contributions of the poles at 
$X=-m_1$ and $X=-m_2$.  So in the end we find that
\begin{equation}
a_{N-1} = \at_{N} = - \sum_{j=1}^{N-1} \at_j .
\label{anminus1}
\end{equation}
Consequently the period $a_{N-1}$ vanishes linearly
with $m$.

Recall that the BPS mass formula 
\cite{Klemm:1995qs,Argyres:1995xh,Hanany:1995na} 
is given by
\begin{equation}
M^2 = 2 |Z|^2~,
\end{equation}
where the central charge $Z$ is 
\begin{equation}
Z = \sum_{i=1}^{N} (n^{i}_e a_i + n^{i}_m a_{D_i})
+ \sum_{j=1}^{2} S_i m_i.
\end{equation}
The constants $n^{i}_e$ and $n^{i}_m$ are the electric
and magnetic charges respectively and satisfy the constraints
$\sum_{i=1}^{N} n^{i}_e = 0 = \sum_{i=1}^{N} n^{i}_m$.
The $S_i$ are the $U(1)$ charges corresponding to global
symmetries that are broken by non-zero values of the
masses.  Also the extra cycles $a_N$ and $a_{D_N}$ are
defined by the conditions $\sum_{i=1}^N a_i = 0$ and
$\sum_{i=1}^N a_{D_i} = 0$ respectively.  The massive
vector bosons then carry electric charges 
$(0,...,0,1,0,...,0,-1,0,...,0)$
while the fundamental matter, quarks, carry
electric charges $(0,...,0,1,0,...,0)
-(1/N,...,1/N)$.  It follows from the expression
for the $a_{N-1}$ period (\ref{anminus1}) that
its associated quark state is becoming massless
in the $m_1, m_2 \rightarrow 0$ limit.
Therefore a hypermultiplet is becoming massless,
or in the dual string theory description, the
probe D7-brane inside the enhancon is really
a  bound state of the D7 and the D3-branes. 
To learn more about this D7-D3 bound state, and
more about the geometry inside the enhancon, we now 
compute the period matrix.

To compute the period matrix we must evaluate the periods
\begin{equation}
\label{periodderivs}
{\partial a_j \over \partial \xi_n} = \oint_{\beta_j} 
\omega_n, \qquad
{\partial a_{D{m}} \over \partial \xi_n} = 
\oint_{\alpha_m} \omega_n~,
\end{equation}
where we define the basis of holomorphic 1-forms by
\begin{equation}
\label{basis}
\omega_n \equiv  - \frac{1}{2 \pi i} \frac{X^{n-1} dX}{Y}, ~
1 \leq n \leq (N-1).
\end{equation}
To evaluate the contour integrals along the
enhancon circle we again use the parametrization given
in (\ref{Xparam}).
To leading order the periods are then given by
\begin{eqnarray}
{\partial \at_{j} \over \partial \xit_n} & = & - \Lambda^{n-N}
\frac{F(n)}{\pi (N-1)} e^{i 2 \pi (j-1)(n-1)/(N-1)} + \cdots,
~~ 1 \leq j \leq (N-1) 
\nonumber \\
{\partial a_{D_{k}} \over \partial \xit_n} & = &
{\partial \at_{j} \over \partial \xit_n}|_{j \rightarrow (k+1/2)},
~~ 1 \leq k \leq (N-2).
\end{eqnarray}

The remaining periods are given by the integrals
\begin{equation}
{\partial a_{D_{N-1}} \over \partial \xit_n}
= - \frac{i}{\pi} \int^{X_{m_1}}_{X_1} \frac{X^{n-1} dX}
{\sqrt{X^{2N} - \Lambda^{2N-2} (X+m_1)(X+m_2)}}. 
\end{equation}
As above we may expand in powers of $m_1$ and $m_2$ to obtain
to leading order the integral
\begin{equation}
{\partial a_{D_{N-1}} \over \partial \xit_n}
= - \frac{i}{\pi} \int^{0}_{\Lambda} \frac{X^{n-1} dX}
{\sqrt{X^{2N} - \Lambda^{2N-2} X^2}} + {\cal O}(m/\Lambda). 
\end{equation}
Redefining the variable of integration by $X=\Lambda e^{x/(N-1)}$,
these periods reduce to 
\begin{eqnarray}
{\partial a_{D_{N-1}} \over \partial \xit_n} & = &
\frac{i}{\pi (N-1)} \Lambda^{n-N} \int^{0}_{- \infty}
dx \frac{e^{x(n - (N+1)/2)/(N-1)}}{\sqrt{2 \sinh (x)}} 
+ {\cal O}(m/\Lambda) \nonumber \\
& = & \frac{\Lambda^{n-N}}{\sqrt{\pi}} \frac{1}{n-1}
\frac{\Gamma(1 + (n-1)/(2N-2))}{\Gamma(1/2 + (n-1)/(2N-2))}
+ {\cal O}(m/\Lambda),
\end{eqnarray}
valid for $n \geq 2$.

The power series expansion in $m_1$ and $m_2$ is not valid
for the $n=1$ case.  In this case one notes that the
factor of $\Xt^{2N}$ in the denominator (defining the
dimensionless variable $\Xt \equiv  X/\Lambda$) can be 
dropped to leading order since the limits of integration
(to leading order) are 0 and 1.  In fact this same approximation
can be used more generally for $n \ll N$ to obtain the
same expression given above.  The resulting integral
is straightforward to evaluate and we find
\begin{eqnarray}
{\partial a_{D_{N-1}} \over \partial \xit_1}  \approx 
- \frac{\Lambda^{1-N}}{\pi} 
\ln \left( \frac{m_1}{4 \Lambda} \zeta - \frac{1}{2} \sqrt{\left( 
\frac{m_1 - m_2}{2 \Lambda} + \frac{m_1}{2 \Lambda} \zeta \right)
\left( \frac{m_2 - m_1}{2 \Lambda} + \frac{m_1}{2 \Lambda} \zeta
\right)} \right),
\end{eqnarray}
where we have defined
\begin{equation}
\zeta \equiv  \sqrt{\left( 1 - \frac{m_2}{m_1} \right)^2 + 4 \left(
\frac{m_2}{\Lambda} \right)^{2(N-1)} }.
\end{equation}
Some special limiting values for
$\partial a_{D_{N-1}} / \partial \xi_1$ are given by
\begin{eqnarray}
{\partial a_{D_{N-1}} \over \partial \xit_1} & \approx &
- \frac{\Lambda^{1-N}}{\pi} \ln \left( \frac{m_2 - m_1}{4 \Lambda}
\right), ~~ \left( \frac{m_1}{\Lambda} \right)^{2(N-1)}
\ll \left( 1 - \frac{m_2}{m_1} \right)^2 \nonumber \\
& \approx & - \frac{\Lambda^{1-N}}{\pi} N \ln \left(
\frac{m_1}{\Lambda} \right), ~~ m_1 = m_2.
\end{eqnarray}

The period matrix, which is expressed in terms of the
periods by
\begin{eqnarray}
\tau_{mj} = \frac{\partial a_{D_m}}{\partial a_{j}}
=  \sum_{n=1}^{N-1} \frac{\partial a_{D_m}}{\partial \xi_n}
\left(\frac{\partial a}{\partial \xi} \right)^{-1}_{nj},
\end{eqnarray}
is now straightforward to compute and is
\begin{eqnarray}
\label{taumatrix}
\tau_{mj} & = & {i \over (N-1)} 
\left(\cot \left( \pi {m-j+1/2 \over N-1} \right)
- \cot \left(\pi {m-j-1/2 \over N-1} \right) \right), ~~ 1 \leq m,j \leq (N-2)
\nonumber \\
& = & {-i \over (N-1)} {e^{-i \pi (m-1/2)/(N-1)} \over
 \sin(\pi (m-1/2)/(N-1))} , ~~  
1 \leq m \leq (N-2),~ j=(N-1) \nonumber \\
& = & i \Lambda^{N-1} {\partial a_{D_{N-1}} \over \partial \xit_1}, ~~
m=j=(N-1),
\end{eqnarray}
where the remaining elements are fixed by symmetry,
$\tau_{mj} = \tau_{jm}$.

This is our final result for the period matrix and one of the main results of
this paper. Let us now see how to interpret this result. 
Near the singularity $m_1=m_2=0$ we can use (\ref{anminus1}) to
express $\tau_{N-1,N-1}$ in terms of $a_{N-1}$
\begin{equation}
\label{insidee}
\tau_{N-1,N-1} \sim - {i \over \pi} \ln ( a_{N-1} + m)
\end{equation}
in the case $m_1 = m_2 = 2$.
This is exactly the singularity we expect when the low energy dynamics
reduces to supersymmetric
QED with two massless hypermultiplets. We interpret this as a bound state
of a D7-brane with the two D3-branes that is able to probe inside the
enhancon shell. We will use the results of this section to compute the
probe metric in section \ref{probesec}.

\section{The single hypermultiplet case}

We consider in this section the technically more
complicated case of the addition of a single
hypermultiplet.  The Seiberg-Witten curve becomes
\begin{equation}
\label{hypercurve}
Y^2 = X^{2N} - \Lambda^{2N-1} (X + m)
\end{equation}
where as before we have taken the vevs of the adjoint scalar
to be zero, i.e., $\phi_j = 0$.  The parameter $m$
is the mass of the hypermultiplet and moreover is
the free parameter that we will vary.  In the D-brane
picture it corresponds to the mass of a string stretched
between the D3-brane probe and the D7-branes.

We consider the case when the D3-brane probe
is close to the background D7-branes, i.e., when $m/\Lambda \ll 1$.
In this limit the roots to the Seiberg-Witten curve are
straightforward to compute in an expansion in powers
of $m/\Lambda$.  In particular one class of roots comes
from taking $m=0$ to leading order to obtain the
equation $X^{2N-1} = \Lambda^{2N-1}$.  These roots
are given by (including some higher order corrections)
\begin{eqnarray}
\label{enhroots}
X_{k} & = & X_{k}^{(0)} (1 + X_{k}^{(1)} + X_{k}^{(2)}+ \cdots) \\
X_{k}^{(0)} & = & \Lambda e^{i 2 \pi k/(2N-1)} \nonumber \\
X_{k}^{(1)} & = & \frac{m}{\Lambda}
\frac{\Lambda^{2N}}{ X_{k}^{(0)} (2 N (X_{k}^{(0)})^{2N-1}
- \Lambda^{2N-1})} \nonumber \\
X_{k}^{(2)} & = & - N (2N-1) \frac{(X_{k}^{(1)})^2 
(X_{k}^{(0)})^{2N-1}}{2 N (X_{k}^{(0)})^{2N-1}
- \Lambda^{2N-1}} \nonumber
\end{eqnarray}
where $1 \leq k \leq (2N-1)$.  
The remaining root comes
from assuming that $X = - m$ to leading order.  Including
some correction terms we find
\begin{eqnarray}
\label{proberoot}
X_{2N} = -m \left( 1 - (\frac{m}{\Lambda})^{2N-1} 
+ \cdots \right).
\end{eqnarray}

\begin{figure}
\centering
\includegraphics[height=7cm]{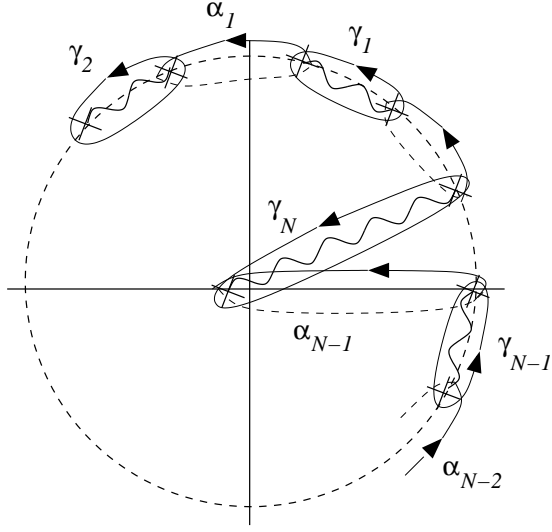}
\caption{Diagram of the branch points and cycles for the
single hypermultiplet case.}
\label{pacman}
\end{figure}

To compute the periods $a_{D_k}$ and $a_j$ we define
the associated cycles
$\alpha_k$ and $\beta_j$ as follows.
The $\alpha_k$ cycle is defined to be the contour encircling
the $X_{2k+1}$ and $X_{2k+2}$ branch points for
$1 \leq k \leq (N-1)$.  The $\beta_j$
cycles are defined in terms of the $\gamma_k$ cycles via
\begin{equation}
\beta_j \equiv  \sum_{k=1}^{j} \gamma_k
\label{gammacycles}
\end{equation}
with the $\gamma_k$ cycle given by the contour that encircles
the $X_{2k}$ and $X_{2k+1}$ branch points for $1 \leq k \leq (N-1)$.
Figure (\ref{pacman})  illustrates the ``pac-man'' nature of the
the choice of branch cuts and cycles 
described above.  We shall denote the periods corresponding
to the $\gamma_k$ cycles by $\at_{k}$.

As in the previous case, we evaluate all periods except for
$\at_{N-1}$ by parameterizing the $X$ integration variable
as
\begin{eqnarray}
\label{param}
 X(\varphi) = X_{k} |_{k \rightarrow k + \varphi/\pi}
\end{eqnarray}
where the limits of integration for $\varphi$ are now just 0 and $\pi$
and $k$ would be $2m+1$ for the $\alpha_m$ cycle and
$2j$ for the $\gamma_j$ cycle.
The main advantage to this parametrization is that
$Y$ reduces exactly to
\begin{equation}
\label{Yparam}
Y = \Lambda^N e^{i \pi k/(2N-1)} e^{i \varphi/(2N-1)} 
\sqrt{(e^{i2 \varphi} -1)}.
\end{equation} 

To evaluate the spectrum of the theory we need to
compute the period integrals (\ref{periodsdefine}),
with a similar expression for $\at_j$ with the 
corresponding contour $\gamma_j$.  The Seiberg-Witten
form $d\lambda$ for the curve (\ref{hypercurve}) is given by
\begin{equation}
d\lambda = - \frac{X^{N+1} dX}{i 2 \pi Y} \biggl(\frac{1}{2(X+m)}
- \frac{N}{X} \biggr).
\end{equation}
Inserting the parametrization (\ref{param}) we find the
following expansion in powers of $(m/\Lambda)$ for
the periods:
\begin{eqnarray}
\at_j & = &  \frac{\Lambda}{4 \pi} \Bigl( e^{i4 \pi j/(2N-1)}
G(N+1) + \frac{m}{(2N-1) \Lambda} G(N) + {\cal O}((m/\Lambda)^2) \Bigr)
\nonumber \\
a_{D_k} & = & \at_{j}|_{j \rightarrow k+ 1/2},
\end{eqnarray}
valid for $1 \leq j \leq (N-1)$ and $1 \leq k \leq (N-2)$.
The coefficients $G(n)$ are given by
\begin{eqnarray}
G(n) & := & 4 \int_{0}^{\pi} d \varphi 
\frac{e^{i \varphi (2n-1)/(2N-1)}}{\sqrt{e^{i2 \varphi}-1}}
\nonumber \\
& = & -i 4 \pi^{3/2} \frac{e^{i \pi (n-1/2)/(2N-1)}}
{\Gamma(\frac{1}{2} + \frac{n-1/2}{2N-1})
\Gamma(1 - \frac{n-1/2}{2N-1})}.
\end{eqnarray}

The remaining period $a_{D_{N-1}}$ is given by the integral
\begin{equation}
a_{D_{N-1}} = \frac{1}{i \pi} \int_{X_{2N}}^{X_{2N-1}}
\frac{X^{N+1} dX}{Y} \biggl(\frac{1}{2(X+m)}
- \frac{N}{X} \biggr).
\end{equation}
The leading order contribution is straightforward to
obtain, one simply sets $m=0$ and reduces the integral
to
\begin{eqnarray}
a_{D_{N-1}} & = &  i \frac{N-1/2}{\pi} \int_{0}^{\Lambda}
\frac{X^{N-1/2} dX}{\sqrt{X^{2N} - 1}} \nonumber \\
& = &  \frac{\Lambda}{\sqrt{2} \pi} \int_{0}^{\infty}
dx \frac{e^{-(1/2 + 1/(N-1/2))x}}{\sqrt{\sinh (x)}}
\nonumber \\
& = &  \frac{\Lambda}{2 \sqrt{\pi}} 
\frac{\Gamma(\frac{1}{2} + \frac{1}{2N-1})}
{\Gamma(1 + \frac{1}{2N-1})}
\end{eqnarray}
where in the second line we have used the change
of variables $(X/\Lambda)^{N-1/2} := e^x$.
A consistency check on the periods arises by noting that
the $a_{D_k}$'s must satisfy 
\begin{equation}
\sum_{k=0}^{N-1} a_{D_k} = m.
\label{constraint}
\end{equation}
This constraint arises by noting that the sum of
contours $\sum_{k=0}^{N-1} \alpha_k$ can be deformed
to a contour encircling the pole of $d\lambda$ at
$X = -m$ and the pole at infinity.  
The residue of $d\lambda$ at both poles
is simply $m/2$ leading to the right-hand-side of
(\ref{constraint}).  One can indeed show that
the leading order piece to the periods of ${\cal O}(\Lambda)$
sums to zero, as it must.  At ${\cal O}(m)$ one
finds that
\begin{equation}
\sum_{k=0}^{N-2} a_{D_k} =  \frac{N-1}{2N-1} m.
\end{equation}
This implies that the ${\cal O}(m)$ contribution
to the $a_{D_{N-1}}$ period must be
$m(1/2 + 1/(2(2N-1))$.  A more careful evaluation
of this period indeed shows that this is the case.
Moreover we have checked that the sum in 
(\ref{constraint}) cancels at ${\cal O}(m^2)$ and
${\cal O}(m^3)$ to ${\cal O}(1/N)$ and ${\cal O}(1)$
respectively.  The period $a_{D_{N-1}}$ is given
to ${\cal O}(m^3)$ 
in appendix B.

The derivatives of the periods obtained in a similar
manner, and are given by
\begin{eqnarray}
\label{periodsnear}
\frac{\partial \at_j}{\partial \xi_n} & = & - \frac{\Lambda^{n-N}}
{2 \pi (2N-1)} e^{i 2 \pi j(n-1/2)/(N-1/2)} \Bigl( G(n) - 
\frac{m}{\Lambda} \Bigl(1 - \frac{n-1}{2N} \Bigr) 
e^{-i 2 \pi j/(N-1/2)} G(n-2N)
\nonumber \\ & + & \cdots \Bigr), 
~~ 1 \leq j \leq (N-1) \nonumber \\
\frac{\partial a_{D_k}}{\partial \xi_n} & = &
\frac{\partial \at_j}{\partial \xi_n} |_{j \rightarrow (k+1/2)},
~ 1 \leq k \leq (N-2)
\end{eqnarray}
where both expressions are valid for $1 \leq n \leq (N-1)$.
The remaining periods are given to leading order by
\begin{equation}
\frac{\partial a_{D_{N-1}}}{\partial \xi_n}
= - \frac{\Lambda^{n-N}}{n \sqrt{\pi}}
\frac{\Gamma(1 + \frac{n}{2N-1})}
{\Gamma(\frac{1}{2} + \frac{n}{2N-1})}.
\end{equation}
We have not been able to find an explicit analytic
form for the period matrix given the formulae above.
While in principle constructing the leading order
piece of the period matrix should proceed as in 
the two-hypermultiplet case, we have not been able
to invert the matrix $\partial \at_k/\partial \xi_n$
analytically.  

\section{Probe metric}
\label{probesec}

To compute the probe metric we should in principle consider the full
$SU(N)\times U(n_3)$ gauge theory dynamics. Since we wish to treat the
D3-branes as probe branes, we will take a limit where the dynamics of
the $U(n_3)$ is frozen out, and the theory reduces to $SU(N)$ coupled
to $n_3$ hypermultiplets. 
 Nevertheless, we obtain the probe metric
from the prepotential of this theory by treating the hypermultiplet
masses $m_f$ as vacuum expectation values of scalars in vector multiplets on the
same footing as the $a_i$. The probe metric is then
\begin{equation}
ds^2_{probe} = {d^2 \CF \over dm^2} dm^2~,
\end{equation}
where the derivative of $\CF$ is defined as
\begin{equation}
\label{probemet}
d\CF = { \partial \CF \over \partial m} dm + \sum_{i=1}^N{\partial \CF \over
  \partial a_i} da_i~.
\end{equation}

Fortunately, D'Hoker et al. \cite{D'Hoker:1997ph} 
have obtained a closed form expression for
the prepotential
\begin{equation}
\label{prepot}
2 \CF = \sum_{i=1}^{N-1} a_i a_{Di} +\sum_{f=1}^{n_3} m_{Df} m_f
+ \frac{1}{2 \pi i} (
Res_{P_+} (z d\lambda) Res_{P_+} (z^{-1} d \lambda) +
Res_{P_-} (z d\lambda) Res_{P_-} (z^{-1} d \lambda))
,
\end{equation}
where 
\begin{equation}
m_{Df} = \int_{P_f}^{P_-} d \lambda~,
\end{equation}
and $P_f$ are the positions of the poles in $d\lambda$, and $P_\pm$ is
infinity on the sheet where $y=\pm \sqrt{A^2-B}$.  The notation
$Res_{P} (z d\lambda)$ denotes the residue of the form $z d\lambda$
at the pole $P$.  The form $d \lambda$ and the function $z$
will be defined below.

To make these $m_D$
integrals well defined, a regularization procedure is needed. To set
this up in a coordinate invariant way, the abelian integral
$E=\log(Y+A)$ is introduced, with\footnote{This $d \lambda$
agrees with our previous expression in (\ref{SWform}) up to
the additive piece $X dX B'/(2B)$ which does not contribute
to the $a_j$ and $a_{D_k}$ integrals.  Rather this term
only contributes to the $m_D$ integrals, hence we do not
change notation here.}
$d \lambda = X dE$. The integrals of
$d\lambda$ are then uniquely defined by imposing the following
asymptotic conditions: near the $P_f$, $z=e^{-E}$,
\begin{equation}
\lambda(z) = -m_f \log z + \lambda(P_f) + \CO(z)~,
\end{equation}
near $P_-$, $z=e^{E \over N-n_3}$,
\begin{equation}
\lambda(z) = -{\rm Res}_{P_-}(z d\lambda)  + \CO(z)
\end{equation}
and near $P_+$, $z= e^{-{E \over N}}$,
\begin{equation}
\lambda(z) =   -{\rm Res}_{P_+}(z d\lambda) + \CO(z)~.
\end{equation}

Computing the prepotential using (\ref{prepot}) is now straightforward
in principle given the results of the previous sections.  In
practice however evaluating the necessary integrals to the
relevant order is tedious.  The period integrals along the
enhancon circle were given in the previous sections to 
linear order in the hypermultiplet masses.  To obtain higher
order corrections one need only expand the integrand to
higher order in the masses.  The integrals can then be done
exactly as discussed for the lower order contributions.
The primary source of difficulty is the $a_{D_{N-1}}$ and
$m_D$ integrals (in either the single or double hypermultiplet
case).  We sketch the computation of these integrals in the
appendix and give their results to ${\cal O}(m^3)$.
Before discussing the limit of $m \ll \Lambda$ relevant
for probing the enhancon, we note that
in the perturbative limit $m\gg \Lambda$,
we recover the supergravity result
(\ref{biaction3}) provided we make the identifications
$\rho_7=\Lambda$. If we also wish the probe D7-brane computation
\cite{Alberghi:2002tu}
match with the supergravity result, we must identify
$\rho_3=\Lambda$. Subleading $1/N$ corrections also appear, with a
form qualitatively the same as the D7-brane probe case described in
\cite{Alberghi:2002tu}. 

In the limit $m\ll \Lambda$, we find
\begin{equation}
\label{fprobemet}
ds^2_{probe} = \CO(N m^0) + \CO(N^0 {m\over \Lambda}) + \CO(N^0 ({m\over
  \Lambda})^2)
+\cdots
\end{equation}
and all terms appear as an
expansion in $m/\Lambda$ with order one coefficients.
The $\CO(N m^0)$ comes purely from the second residue
term appearing in the formula for the prepotential (\ref{prepot}).
This term however is necessary in order to recover the
correct perturbative form of the prepotential.

Computing the curvature of the probe metric in the form of the 
Ricci scalar, we find
\begin{equation}
R\sim {\Lambda^2\over N (\log(|m|/\Lambda))^3 |m|^2}
\end{equation}
when $|m|\gg \Lambda$. At a typical point, the curvature is of order
$1/N$, and becomes of order one near the enhancon shell. 

When $m\ll \Lambda$
\begin{equation}
R\sim \CO (N^{-2} m^0) +\cdots
\end{equation}
This is the one of the main results of this paper. This result 
is consistent with the proposed flat interior for the 
enhancon shell of \cite{Johnson:1999qt} as it does reduce
to flat geometry as $N \to \infty$.  In obtaining this
result it was crucial to include the residue terms in
the formula for the prepotential (\ref{prepot}).  Without
these terms the $\CO (m^0)$ piece of the metric (\ref{fprobemet})
would have come with an $\CO (1)$ coefficient, so that the
curvature would also have been $\CO (1)$.

In \cite{Alberghi:2002tu} we considered a D7 probe near the enhancon
shell, where critical behavior emerges. This limit has also been
studied by Ferrari \cite{Ferrari:2001mg,Ferrari:2001zb}. It would be
interesting to study the analogous limit for the D3 probe, but we
leave that for the future.

\section{Conclusions}

In this paper we have explored the enhancon mechanism from the point
of view of $\CN=2$ supersymmetric $SU(N)$ gauge theory with
fundamental hypermultiplets. We have
obtained the exact low energy effective action for this theory,
including non-perturbative effects. 
Our earlier results \cite{Alberghi:2002tu}
explored the moduli space metric of the pure $SU(N)$ gauge
theory, including configurations with probe branes both far and near
the enhancon shell. In the present work, 
we studied the enhancon in the presence of D3
brane probes which allowed us to probe the region inside the enhancon shell.
In particular, we were able to show that at large
$N$ the geometry inside the enhancon shell is flat, i.e., curvature
appears at order $1/N^2$.
Our results support the
supergravity picture of the enhancon resolution mechanism 
advocated in \cite{Johnson:1999qt}, and improve
upon it in the gauge theory limit by including the full dynamics of
the brane sources.  

The gauge theory description makes clear the special role played by
spherical symmetry in this resolution of the repulson
singularity. The enhancon is at best marginally stable, since there
exist exactly flat directions in moduli space corresponding to the
quantum corrected position coordinates $a_i$ of the D7-branes.
{}From the gauge theory perspective, one can easily
construct configurations where multiple branch cuts collide in the
curve (\ref{twohypercurve}), though
these points in moduli space will generally be non-spherically
symmetric. These degenerations give rise to extra light particles
creating singularities in the moduli space metric.
Some of these degenerations are related to conformal field
theories. From the supergravity viewpoint, these should be dual to nonsingular
throats opening up in the geometry of the form anti-de Sitter space
cross a sphere. Others are related to Argyres-Douglas points, and
it remains unknown how these singularities will be resolved from the supergravity
viewpoint.
One possibility is there is no satisfactory resolution of these
singularities purely within supergravity, and one must look for a more
intrinsically stringy method of singularity resolution \cite{Lowe:2000ir}.

\bigskip
\centerline{\bf Acknowledgements}
\noindent
We thank G.L. Alberghi for collaboration during the early stages of
this work. We thank E. D'Hoker for a helpful comment.
This research is supported in part by DOE
grant DE-FE0291ER40688-Task A.

\appendix
\section{Appendix}

In this appendix we collect some facts concerning
the series expansions of various quantities
in powers of the dimensionless quantities
$m_1/\Lambda$ and $m_2/\Lambda$ for the two hypermultiplet
case studied in (\ref{sec:twohyper}).  To begin with
we consider the roots of the Seiberg-Witten curve,
\begin{equation}
\Yt^2 = \Xt^{2N} - (\Xt + \mt)(\Xt + \xi \mt),
\end{equation} 
rewritten in terms of dimensionless variables
$\Yt \equiv  Y/\Lambda^{2N}$, $\Xt \equiv  X/\Lambda$,
$\mt \equiv  m/\Lambda$, and $\xi = m_2/m_1$ 
with $m = m_1$.  We search for roots of the
polynomial by demanding that solutions of the form
\begin{equation}
\Xt = \Xt^{(0)} (1 +  \sum_{j=1}^{\infty} \mt^{j} \Xt^{(j)})
\end{equation}
hold order by order in the expansion parameter $\mt$.
To zeroth order this will yield the solutions
found previously in (\ref{twoenhroots}) for
$X^{(0)}_k$.  The coefficient of the $\mt^{j}$
term in the polynomial will then yield a linear
equation for $\Xt^{(j)}$ in terms of
the $\Xt^{(l)}$'s for $0 \leq l \leq (j-1)$.
Consider in particular the case where
$\Xt^{(0)} = e^{i \pi k/(N-1)}$.  It follows by
induction that $\Xt^{(j)} \propto e^{- i \pi j k/(N-1)}$
where the coefficient of proportionality is independent
of $k$.  Denoting this coefficient by 
$d_j(\xi)$ we find the expansion for the roots
\begin{equation}
\Xt_k = e^{i \pi k/(N-1)} \sum_{j=0}^{\infty} d_j (\xi) \mt^{j} 
e^{- i \pi j k/(N-1)}.
\label{rootexp}
\end{equation}

Now we can use this expansion to prove the statement
in (\ref{sec:twohyper}) that the series expansion
of $\Yt$ (working with dimensionless quantities now) 
is given simply by (\ref{Yexpans}).  Recall
that we substitute for $\Xt$ in $\Yt$ the expansion
$\Xt_k|_{k \rightarrow k+\varphi/\pi}$.  Expanding
$\Yt$ in powers of $\mt$, one obtains exactly the
same coefficients for each power $\mt^{j}$
as above when solving for the roots, providing
that these coefficients are expressed in terms
of the $\Xt^{(j)}$'s and not their explicit
expressions.  The only exception to this occurs
for $j=0$.  It follows immediately that
the coefficients of the $\mt^{j}$ terms vanish
by construction for $j \geq 1$.   On the other
hand the $j=0$ term yields precisely (\ref{Yexpans}).

A simple consequence of the expansion of the roots (\ref{rootexp})
and the parametrization (\ref{Xparam}) is the
expansion of the periods given in (\ref{periodseries}).
This follows easily by inserting the expansion
(\ref{rootexp}) evaluated at $k \rightarrow k+\varphi/\pi$
into the Seiberg-Witten period $d\lambda$ and using
the expansion of $Y$ just derived.

\section{Appendix}

In this appendix we present the results of the $a_{D_{N-1}}$
and $m_{D}^{bare}$ period computations in both the single and double
hypermultiplet cases and sketch briefly how these expressions
were derived.  For the single hypermultiplet case we find (setting
$\Lambda = 1$ here and in the remaining formulae in this appendix)
\begin{eqnarray}
a_{D_{N-1}} & = &  \frac{1}{2 \sqrt{\pi}} \frac{\Gamma (\frac{1}{2} + 
\frac{1}{2N-1})}{\Gamma (1 +  \frac{1}{2N-1})} + \frac{m}{2} \Bigl(1
+ \frac{1}{2N-1} \Bigr) \\ \nonumber
& + &  \frac{m^2}{4 \sqrt{\pi} (2N-1)}
\frac{ \Gamma (\frac{1}{2} - \frac{1}{2N-1}) }
{  \Gamma (1 - \frac{1}{2N-1}) }
\Bigl( 1 + \frac{1}{2N-1} \Bigr) \\ \nonumber
& - & \frac{m^3}{6 \sqrt{\pi} (2N-1)}
\frac{ \Gamma (\frac{1}{2} - \frac{2}{2N-1})} 
{\Gamma (1 - \frac{2}{2N-1})}
\Bigl( 1 - \frac{9}{2N-1} + \frac{2}{(2N-1)^2} \Bigr)
\\ \nonumber
& \approx &  \frac{1}{2} + \frac{m}{2} \bigl((-1)^N
+ \frac{1}{2N-1} \bigr)
 + \frac{m^2}{8 N}
- \frac{m^3}{12 N}
\end{eqnarray}
and
\begin{eqnarray}
m_{D}^{bare} & = & \frac{N-1}{i 2 \pi} X_{\infty} + \frac{m}{i 2 \pi} 
\ln (X_{\infty}) - \frac{m \ln (\delta m/m^N)}{i 2 \pi}
+ m \biggl( \frac{\ln (2) +1}{i 2 \pi} - \frac{(i 2 \ln (2)
- \pi)}{4 \pi (2N-1)} \biggr) \\ \nonumber
& - & 
\frac{(-1)^{1/2 - 1/(2N-1)}}{8 \pi^{3/2}} m^2 \Gamma \Bigl( \frac{1}{2}
- \frac{1}{2N-1} \Bigr) 
\biggl( \Gamma \Bigl(2 + \frac{1}{2N-1} \Bigr) + \frac{2}{2N-1}
\Bigl( \Gamma \Bigl(1 + \frac{1}{2N-1} \Bigr) \\ \nonumber
& + &
\Gamma \Bigl(\frac{1}{2N-1} \Bigr) \Bigr) \biggr)
 + 
\frac{(-1)^{1/2 - 2/(2N-1)}}{4 \pi^{3/2}} m^3 \Gamma \Bigl( \frac{1}{2}
- \frac{2}{2N-1} \Bigr) 
\biggl( \frac{1}{6}  \Gamma \Bigl(3 + \frac{2}{2N-1} \Bigr) \\
\nonumber
& + & \frac{1}{2N-1}
\frac{1}{2} \Bigl( \Gamma \Bigl(2 + \frac{2}{2N-1} \Bigr) +
\Gamma \Bigl(1 + \frac{2}{2N-1} \Bigr) +
\Gamma \Bigl(\frac{2}{2N-1} \Bigr)
\Bigr) \biggr) \\ \nonumber
& \approx & \frac{N-1}{i 2 \pi} X_{\infty} + \frac{m}{i 2 \pi} 
\ln (X_{\infty}) - \frac{m \ln (\delta m/m^N)}{i 2 \pi}
+ m \frac{\ln (2) +1}{i 2 \pi} -i \frac{3}{8 \pi} m^2
+i \frac{5}{24 \pi} m^3,
\end{eqnarray}
where we have given both the exact expression to cubic order in $m$
and the large $N$ approximate form.
Similarly for the double hypermultiplet case we find
\begin{eqnarray}
a_{D_{N-1}} & = &  \frac{N-1}{N \sqrt{\pi}} \frac{\Gamma (\frac{3}{2} + 
\frac{1}{2(N-1)})}{\Gamma (1 +  \frac{1}{2(N-1)})}
+ \frac{m^2}{\sqrt{\pi} (N-2)} \Gamma \Bigl(\frac{3}{2} -
\frac{1}{2(N-1)} \Bigr) \biggl(\frac{2(N-1)}{\Gamma(-1 -
  \frac{1}{2(N-1)})}
\\ \nonumber
& + &
\frac{N+1}{\Gamma (-\frac{1}{2(N-1)})} -
\frac{1}{\Gamma(1 -  \frac{1}{2(N-1)})} \biggr) \\ \nonumber
& + &  \frac{m^3}{\sqrt{\pi} (N-3)} \Gamma \Bigl(\frac{3}{2} -
\frac{1}{N-1} \Bigr) \biggl( \frac{8(N-1)}{3 \, \Gamma(-2  - \frac{1}{N-1})}
+ \frac{2(2N+1)}{\Gamma(-1  - \frac{1}{N-1})} \\ \nonumber
& - & \frac{1}{\Gamma( - \frac{1}{N-1})}
+ \frac{1}{\Gamma(1 - \frac{1}{1N-1})} \biggr) \\ \nonumber
& \approx & \frac{1}{2} - \frac{m^2}{4 N}
- \frac{m^3}{6 N}
\end{eqnarray}
and
\begin{eqnarray}
m_{D}^{bare} & = & \frac{1}{i 2 \pi} \biggl( -(N-2) X_{\infty}
- 2 m \ln (X_{\infty}) + 2 m \ln(\delta m/m^{N/2}) \\ \nonumber
& - & \frac{(-1)^{1/(2(N-1))}}{\sqrt{\pi}} \frac{N-1}{N}
\Gamma \Bigl( \frac{3}{2} + \frac{1}{2(N-1)} \Bigr)
\Gamma \Bigl(- \frac{1}{2(N-1)} \Bigr) +
m \Bigl(2 - 2 \ln (2) + \frac{i \pi}{2} \Bigr) \\ \nonumber
& - & m^2 \frac{(-1)^{1/(2(N-1))}}{2 \sqrt{\pi}}
\frac{N}{(N-1)(N-2)} \Gamma \Bigl( \frac{3}{2} - \frac{1}{2(N-1)} \Bigr)
\Gamma \Bigl(\frac{1}{2(N-1)} \Bigr) \\ \nonumber
& + & m^3
\frac{(-1)^{-1/(N-1)}}{3 \sqrt{\pi}} \frac{N(7N-11)}{(N-1)^2 (N-3)}
\Gamma \Bigl( \frac{3}{2} - \frac{1}{N-1} \Bigr)
\Gamma \Bigl(\frac{1}{N-1} \Bigr) \biggr) \\ \nonumber
& = & \frac{1}{i 2 \pi} \biggl( -(N-2) X_{\infty}
- 2 m \ln (X_{\infty}) + 2 m \ln(\delta m/m^{N/2})
+ N \\ \nonumber 
& + & m \Bigl(2 - 2 \ln (2) + \frac{i \pi}{2} \Bigr)
- \frac{m^2}{2} + \frac{7 m^3}{6} \biggr).
\end{eqnarray}

For concreteness we shall now briefly sketch the derivation
of $a_{D_{N-1}}$ for the single hypermultiplet.  The remaining
periods follow via identical reasoning however.  The integral 
that we need is given by
\begin{equation}
a_{D_{N-1}} = \oint_{\alpha_{N-1}} d\lambda =
\int_{X_0}^{X_{2N}}  d\lambda
\end{equation}
where $d\lambda$ is given by (\ref{SWform}).
To evaluate the integral we shall split the 
integration region into four parts:
(i) $ (X_{2N} + m) \leq (X + m) \leq \xi_0 m^{2N} $,
(ii) $ \xi_0 m^{2N}\leq (X + m)\leq \xi_{1} m$, 
(iii)$ \xi_{1} m\leq (X + m)\leq (X_0 +m - \xi_{2} m/N)$,
and (iv) $ (X_0 - \xi_{2} m/N) \leq X \leq X_0$.
The parameters $\xi_{j}$ are all numbers of ${\cal O}(10)$.
In region (i) we make the change of integration
variable to $X = -m + u m^{2N}$.  The integrand can
then be series expanded in $m$ and the resulting 
integrals are straightforward to evaluate (at least
to the order presented above).  
In region (ii) one series expands the integrand
in the variable $(X+m)$. 
In region (iii) one series expands the integrand
in $m$, and in region (iv) one defines the variable
of integration $X = X_0 - v m/N$ and then series
expands in $m$.  It is straightforward to check
that these series expansions are valid over the
ranges given above.  As a consistency check one
notes that the full integral must be independent
of the (relatively) arbitrary parameters $\xi_j$
introduced in the intermediate steps.  Indeed
one can show that this cancellation takes place
to the order that we are working.

\bibliography{enhbib}
\bibliographystyle{hunsrt}
\end{document}